\magnification=\magstep1
\baselineskip = 13pt
\parskip=3pt
\def\L{{\bf L}}
\def\S{{\bf S}}
\def\J{{\bf J}}

\def\n{{\bf n}}
\def\tr{{\rm Tr}}
\def\lp{\bigl(}
\def\rp{\bigr)}
\def\>{\rangle}
\def\k#1{|#1\>}
$\vphantom{.}$
\vskip -40pt

\font\title = cmbx10 scaled 1440

\newcount\eqnumber

\def\eq(#1){
    \ifx\DRAFT\undefined\def\DRAFT{0}\fi	%if undef'd, make it 0
    \global\advance\eqnumber by 1%
    \expandafter\xdef\csname !#1\endcsname{\the\eqnumber}%
    \ifnum\number\DRAFT>0%
	\setbox0=\hbox{#1}%
	\wd0=0pt%
	\eqno({\offinterlineskip
	  \vtop{\hbox{\the\eqnumber}\vskip1.5pt\box0}})%
    \else%
	\eqno(\the\eqnumber)%
    \fi%
}
\def\(#1){(\csname !#1\endcsname)}

\centerline{\title On Conway and Kochen's} 
\vskip 5pt

\centerline{\title ``Thou shalt not clone one bit!''}
\vskip 10pt
\centerline{N. David Mermin}
\centerline{Laboratory of Atomic and Solid State Physics}
\centerline{Cornell University, Ithaca, NY 14853-2501}
\vskip 10pt

{

\narrower \narrower

\noindent A physicists' explanation of why thou shalt not clone
one bit gives a stronger version of Conway and Kochen's theorem.

}

\vskip 8pt

John Conway and Simon Kochen prove in arXiv:0711.2310 [quant-ph]  that
it is impossible to have three spins-1, of which two are ``twinned'',
while the third necessarily gives the same answer as those two do to a
``spin-zero measurement'' along one given direction.  A {\it spin-zero
measurement\/} along \n\ is a measurement that asks whether or not the
spin along the direction \n\ is zero.  Two spins-1 are {\it twinned\/}
if each gives the same answer (yes or no) to a spin-zero measurement
along \n, regardless of the choice of \n.

For example, two spin-1 particles in the singlet (zero total angular
momentum state) $$\k\Phi = (1/\sqrt3)\bigl(\k0\k0 - \k1\k{-1} -
\k{-1}\k1\bigr) \eq(0)$$ are twinned in the Conway-Kochen sense,
because $\k\Phi$ has the form \(0) regardless of the common direction
\n\ along which the two spin components are specified.  In this
particular case a stronger form of the Conway-Kochen theorem follows
immediately from the fact that a system in a {\it pure state\/} (in
this case the two twinned spins) can have {\it no correlations
whatever with any external system\/} (in this case the third spin).  A
proof of this elementary but fundamental property of pure states, a
pillar of the ``Ithaca Interpretation of Quantum Mechanics'', can be
found in Appendix B of arXiv:quant-ph/9801057.

It is plausible to conjecture that requiring the twin condition to
hold for {\it arbitrary\/} directions \n\ is so strong a constraint
that it can happen {\it only\/} in the rotationally invariant singlet
state.  The stronger Conway-Kochen theorem would then follow
immediately from the general fact that a system in a pure state can
have no external correlations.  In the remainder of this note I
establish this conjecture with a simple but slightly unconventional
application of elementary angular momentum technology.

Call the angular momentum vector operators (in units of $\hbar$) for the
two spins-1 \L\ and
\S.  The twin condition is the requirement that the measured squared
spin components of the two spins are both 0 or both 1, regardless of the common
direction \n\ along which they are measured.  The analytical expression
of the twin condition is
$$0 = \tr\lp\rho\bigl[(\n\cdot\L)^2-(\n\cdot\S)^2\bigr]^2\rp\eq(1)$$
where $\rho$ is the density matrix for the two twinned spins.  We must
prove that \(1) can hold for arbitrary directions \n\ only if $\rho$
is the projection operator on the singlet state.  
It follows from \(1) that any eigenstate $\k\Psi$ of $\rho$ associated
with a non-zero (necessarily positive) eigenvalue must satisfy
$$  \bigl[(\n\cdot\L)^2-(\n\cdot\S)^2\bigr]\k\Psi = 0.\eq(2)$$
We now prove that the only $\k\Psi$ satisfying \(2) for all
directions \n\ is the singlet state, and therefore $\rho$ must indeed be the
singlet-state projection operator.

Eq.~\(2) is equivalent to the requirement that
$$\lp S_iS_j+S_jS_i\rp\k\Psi =
\lp L_iL_j+L_jL_i\rp\k\Psi 
\eq(3)$$ 
for all the components of \L\ and \S\ along any three orthogonal directions
$i,j = x,y,z.$
Act on both sides of \(3) with $L_iL_j$, sum on $i$ and $j$, use
the fact that $$\L^2 = L_x^2+L_y^2+L_z^2 =2 = \S^2 = S_x^2+S_y^2+S_z^2 \eq(4)$$ and the standard
angular-momentum commutation relations

$$ L_xL_y - L_yL_x = i L_z,\ et\ cyc.,$$ 
\vskip -10pt
$$ S_xS_y - S_yS_x = i S_z,\ et\ cyc.,$$ 
\vskip -10pt
$$ L_iS_j - S_jL_i = 0,\eq(5)$$ 
to get from \(3)
$$\bigl[ 2(\L\cdot\S)^2 + (\L\cdot\S)\bigr]\k\Psi =
\bigl[ 2(\L^2)^2 - (\L^2)\bigr]\k\Psi =
 6\k\Psi.\eq(6)$$

Now note that for two spins-1 the 9-dimensional space of two-spin
states $\k\Psi$ is the direct sum of 5-dimen\-sion\-al ($j=2$), 3
dimensional ($j=1$), and 1-dimensional ($j=0$, singlet) subspaces of
total angular momentum $\J = \L + \S$.  Since in each such subspace
$$j(j+1) = \J^2 = \L^2 + 2\L\cdot\S + \S^2 = 4 + 2\L\cdot\S,
\eq(6a)$$
the states in each of the three subspaces are eigenstates of $\L\cdot\S$ with
eigenvalues $$1\ (j=2),\ -1\ (j=1),\ -2\ (j=0).\eq(7)$$ 

The projection operators on these subspaces commute with $\L\cdot\S$.
Therefore if we act on \(6) with projection operators on the $j=2$,
$j=1$, and $j=0$ subspaces then \(7) requires the projection of
$\k\Psi$ on the $j=2$ and $j=1$ subspaces to vanish, but allows the
projection of $\k\Psi$ on the (one dimensional) $j=0$ subspace not to
vanish.  So the two-spin density matrix $\rho$ for twinned spins must
indeed be a projection operator on the singlet state, and measurement
outcomes on any third spin (or any other external system) must
therefore be statistically independent of any measurement outcomes on
the twins.

%Note that twin behavior along just the 5 directions \x, \y, \x+\y,
%\y+\z, and \z+\x\ is enough to require the state of a pair to be the
%singlet state, and therefore prohibit any third party correlations.
\bye